\documentclass[journal]{IEEEtran}
\usepackage{amsmath,amsfonts}
\usepackage{algorithmic}
\usepackage{algorithm}
\usepackage{array}
\usepackage[caption=false,font=normalsize,labelfont=sf,textfont=sf]{subfig}
\usepackage{xcolor}
\usepackage{textcomp}
\usepackage{stfloats}
\usepackage{url}
\usepackage{verbatim}
\usepackage{graphicx}
\usepackage{cite}
\usepackage{amssymb}
\usepackage{booktabs}
\begin{document}

\title{Asynchronous Random Access in Massive MIMO Systems Facilitated by the Delay-Angle Domain}

\author{Ao Chen, Wei Chen, \IEEEmembership{Senior Member, IEEE}, Bo Ai, \IEEEmembership{Fellow,~IEEE}, Petar Popovski, \IEEEmembership{Fellow,~IEEE}

\thanks{Ao Chen, Wei Chen and Bo Ai are with the School of Electronic and Information Engineering, Beijing Jiaotong University, Beijing 100044, China (e-mail: aochen@bjtu.edu.cn; weich@bjtu.edu.cn; boai@bjtu.edu.cn).}
\thanks{Petar Popovski is with the Department of Electronic Systems, Aalborg University, 9220 Aalborg, Denmark (e-mail: petarp@es.aau.dk).}
}



\maketitle

\begin{abstract}
The problem of uplink transmissions in massive connectivity is commonly dealt with using schemes for grant-free random access. When a large number of devices transmit almost synchronously, the receiver may not be able to resolve the collision. This could be addressed by assigning dedicated pilots to each user, leading to a contention-free random access (CFRA), which suffers from low scalability and efficiency. This paper explores contention-based random access (CBRA) schemes for asynchronous access in massive multiple-input multiple-output (MIMO) systems. The symmetry across the accessing users with the same pilots is broken by leveraging the delay information inherent to asynchronous systems and the angle information from massive MIMO to enhance activity detection (AD) and channel estimation (CE). The problem is formulated as a sparse recovery in the delay-angle domain. The challenge is that the recovery signal exhibits both row-sparse and cluster-sparse structure, with unknown cluster sizes and locations. We address this by a cluster-extended sparse Bayesian learning (CE-SBL) algorithm that introduces a new weighted prior to capture the signal structure and extends the expectation maximization (EM) algorithm for hyperparameter estimation. Simulation results demonstrate the superiority of the proposed method in joint AD and CE.
\end{abstract}


\section{Introduction}
\IEEEPARstart{M}{assive} machine-type communication (mMTC) is one of the three key connectivity types from the fifth generation (5G) wireless communication system and is expected to be further enhanced in the sixth generation (6G) system \cite{shahzadi20216g,chen2023signal}. Typical applications in mMTC include smart cities, environmental monitoring, and event detection, where the main goal is to provide connectivity for massive devices with sporadic data characteristics. Given that user packets are typically small and only a subset of users need to transmit data to the base station (BS) simultaneously, traditional grant-based access mechanisms with their complex signaling interactions prove to be inefficient \cite{bockelmann2016massive}. To address the needs of sporadic traffic and small data packets for numerous users in upcoming wireless communication systems, grant-free random access is widely favored for its simple signaling interaction and low overhead\cite{choi2021grant,8454392}. In grant-free random access, an active user transmits its pilot and packet directly to the BS without requesting prior approval, leading to low signal overhead and access delay, thereby greatly enhancing access efficiency. This capability is essential for supporting the growing number of IoT devices and applications that demand efficient, reliable communication in future networks.

\par
One of the main challenges in grant-free random access is joint sparse activity detection (AD) and channel estimation (CE), which aims to identify the active devices among all users based on the transmitted pilot sequence. A straightforward strategy is contention-free random access (CFRA), in which the BS allocates a unique pilot sequence to each user in advance, and then the user uses this pilot sequence for joint AD and CE. However, since only a few users access the network simultaneously, only a fraction of the pilot sequences are utilized, resulting in inefficient utilization of pilot resources. Additionally, this strategy suffers from poor scalability due to the finite number of pilot sequences, which limits the number of users that can be supported. An alternative strategy, namely contention-based random access (CBRA), uses a common pool of pilot sequences for all users to select in a random manner \cite{kang2022scheduling}. This allows for more efficient use of pilot sequences and better scalability. However, it is inevitable that multiple users might choose the same pilot simultaneously, resulting in pilot collisions. Finding ways to reduce pilot collisions in CBRA while maintaining spectrum efficiency presents a significant challenge. 
\par
Another challenge in grant-free random access is synchronization, which is crucial for efficient data transmission and requires coordination between the BS and users. However, many existing works are based on the assumption of perfect synchronization without considering the extra cost and difficulty. Asynchronous grant-free random access has attracted significant attention in recent years, while most works treat asynchronism as a negative factor and focus on devising techniques to mitigate performance degradation due to asynchronism. On the contrary, asynchronism can be leveraged to break the symmetry among the accessing devices and bring 
additional degrees of freedom in the signal space that can be utilized in grant-free random access. 

This paper presents approaches to use asynchronism towards improving the performance of AD and CE. We consider asynchronous grant-free CBRA in massive multiple input multiple output (MIMO) systems and propose a new approach to deal with the joint AD and CE problem in a mMTC scenario. The proposed approach exploits the time delays caused by asynchronism and the angle information brought by massive MIMO to reduce pilot collisions in CBRA. The problem of joint AD and CE then boils down to a linear inverse problem (LIP) with a joint row-sparse and cluster-sparse structure. To address this problem, we develop a cluster-extended sparse Bayesian learning (CE-SBL) algorithm that exploits the embedded signal structure to obtain improved performance of joint AD and CE for grant-free CBRA in mMTC.  

\subsection{Related Work}
The sporadic traffic of users naturally leads to a sparse communication pattern and this has been often addressed by using compressed sensing (CS) techniques\cite{schepker2011sparse,chen2018sparse,shao2019dimension}. The joint AD and CE in massive random access is closely related to the sparse recovery problem in CS, which exploits the sporadic traffic in MMTC. A typical CS algorithm, orthogonal matching pursuit (OMP), was initially used in multi-user detection\cite{schepker2011sparse}, and it proves the rationality and effectiveness of using CS theory to solve the problem by experiments. The approximate message passing (AMP) is another attractive algorithm for massive random access due to its efficient computational performance. For example, Chen et al. explore two scenarios based on whether the BS knows the large-scale component of channel fading or not to solve the problem of joint AD and CE\cite{chen2018sparse}. They create a denoiser for AMP using minimum mean squared error (MMSE) and utilize channel statistics to solve the problem. Additionally, they present an analytical description of the probabilities of false alarms and miss detection using state evolution. Sparse Bayesian learning (SBL) is also a widely used method for sparse signal recovery. Wipf and Rao have empirically demonstrated the superior performance of SBL in signal recovery\cite{wipf2004sparse} and extended it to the case of multiple responses to solve the sparse recovery problem in\cite{wipf2007empirical}, named M-SBL. This expands the scope of application of the CS algorithm so that it can be used in multi-response systems, such as the MIMO systems. Furthermore, by using the structural features existing in the signal space, we could obtain better recovery performance. For example, Fang et al. propose a variant of SBL named PC-SBL\cite{fang2014pattern}, which uses a pattern-coupled Gaussian prior to capturing the relationship between neighbouring elements and encouraging block sparsity. This method aims to interconnect the sparse patterns of neighbouring coefficients, potentially promoting structured sparse solutions even without prior knowledge of cluster patterns. 


\par
The CBRA is widely used due to its good scalability and low complexity. It originates from slot ALOHA\cite{casini2007contention,liva2010graph}, which divides time into discrete slots and allows devices to transmit at the beginning of each selected slot. If a collision occurs, the transmitting devices are informed and can attempt retransmission at a later slot. Sørensen et al. propose a random access method for the massive MIMO systems called coded pilot random access\cite{8487005}, which is a protocol that enables the basic characteristics of MIMO systems to perform channel estimation and resolve conflicts. To improve scalability, Bai et al. introduce a CBRA scheme that implements a one-step access approach to conserve energy and spectrum resources to enable extensive access within a specific time-frequency resource\cite{bai2020contention}. They also propose a receiver algorithm to obtain the optimal solution to the problem, in which the prior information is used to improve the data recovery performance\cite{bai2020contention2}. In addition, Chen et al. propose an SBL-based algorithm to deal with the CBRA problem, which utilizes the structural features of row sparsity and cluster sparsity of massive MIMO channels in the angular domain to enhance AD and CE performance\cite{chen2021joint}.

\par
All the works mentioned above are based on the assumption of perfect system synchronization. However, achieving synchronization over all devices is very difficult and costly. It is necessary to consider asynchronous scenarios. For example, an extension of the AMP algorithm named orthogonal AMP (OAMP) is considered to be used in asynchronous GFRA systems, which exploits the common sparsity between pilot signals received by multiple BS antennas\cite{10436887}. The authors further propose the memory AMP (MAMP) algorithm to reduce the algorithm complexity\cite{10436887}. The lack of synchronization among users complicates processing on the receiver side, resulting in a more intricate receiver design. To relieve the pressure of the receiver, a selective interference elimination aided orthogonal matching pursuit (SIE-OMP) algorithm is proposed in\cite{10168894}, which reduces the complexity of joint AD and CE with a reduced-size set of candidate pilots. Furthermore, different from the CS-based method, Wang et al. propose a covariance-based method to solve AD in asynchronous random access, which characterizes the problem as a maximum likelihood estimation (MLE) problem, relying on the sample covariance of the received signal\cite{wang2022covariance}. Asynchronous for other important system are also considered and even utilized. For example, the authors in \cite{9916169,10278871} investigate synchronous random access in cell-free MIMO systems and design an accelerated distributed algorithm to overcome the problem of synchronization between devices due to the low cost of local oscillators. Moreover, Torlak and Xu consider the asynchronous code division multiple access (CDMA) systems and leverage the delay information resulting from asynchronism as auxiliary data to improve channel estimation \cite{torlak1997blind}. To overcome and exploit asynchronous information, a probabilistic data association (PDA) method is extended to multiuser detection in asynchronous CDMA communication channels\cite{luo2003sliding}. The simulation results show that this method achieves better performance than synchronous under the premise of slightly increasing complexity by utilizing asynchronous information.

\subsection{Main Contributions}
This paper considers the asynchronous uplink grant-free CBRA in massive MIMO systems. To achieve enhanced massive access, we exploit the additional degrees of freedom brought by the asynchronous scenario and massive MIMO. Specifically, as shown in Fig. \ref{fig_3}, our objective is to resolve conflicts of active users selecting the same pilot, while having different angles of arrival (AOAs) or delays. To take advantage of the characteristics of the structural signal, an SBL-based method, named cluster-extended SBL (CE-SBL), is proposed, which simultaneously exploits row-sparse and cluster-sparse properties of MIMO channels in the asynchronous delay-angle domain. The main contributions of this paper are summarized as follows:
\begin{figure}[!t]
\centering
\includegraphics[width=3.5in]{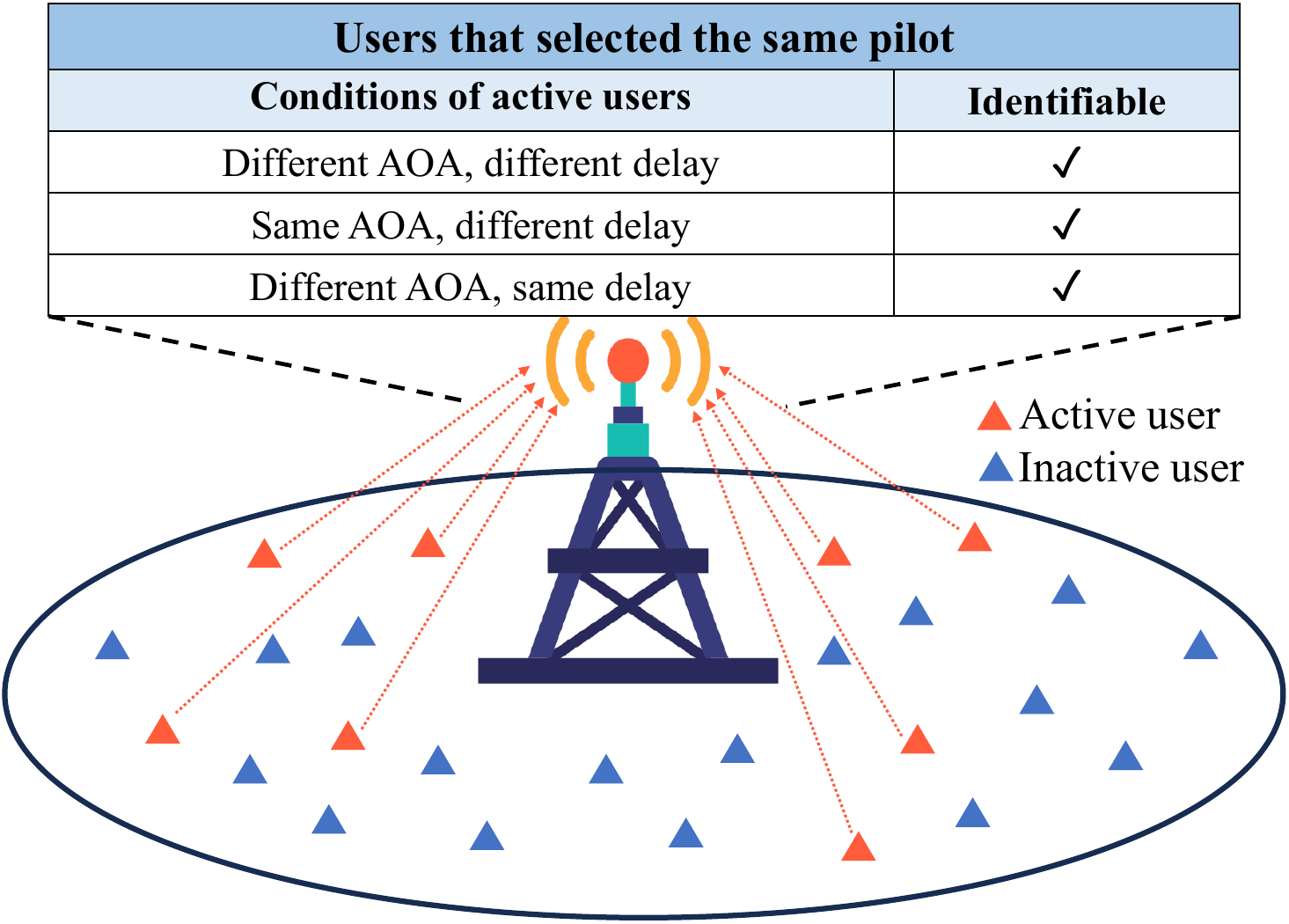}
\caption{Delay-angle domain aided massive connectivity network}
\label{fig_3}
\end{figure}
\begin{itemize}
    \item We propose to exploit delays caused by asynchronism, in addition to AOAs in massive MIMO systems, to resolve conflicts in synchronous uplink grant-free CBRA in massive MIMO systems. We formulate the extended system model in the delay-angle domain for massive MIMO systems. In this model, even when multiple users choose the same pilot and collide, they can be distinguished by different AOAs and delays.
    \item We formulate the multiple measurement vector (MMV) problem of joint AD and CE as an LIP in the delay-angle domain, where MIMO channels have row-sparse and cluster-sparse structures simultaneously. The clusters with non-zero elements may have different sizes and supports, which presents new challenges for algorithm design.
    \item We propose an SBL-based algorithm, which supposes a weighted prior to capture cluster structures and generalize the expectation maximization (EM) algorithm\cite{dempster1977maximum,moon1996expectation} on the new objective function. The algorithm can achieve high-precision sparse signal recovery. We provide theoretical analysis to unveil the superiority of the algorithm for asynchronous massive connectivity. 
\end{itemize}
\subsection{Paper Organization and Notation}
The rest of the paper is organized as follows. In Section II, we describe the system model for the asynchronous grant-free CBRA scenario, and introduce the SBL method for the multi-response model. In Section III, we introduce the CE-SBL algorithm design for asynchronous massive access in massive MIMO systems. Numerical results are presented in Section IV. Finally, we conclude this paper in Section V.
\par
The notations in this paper are defined as follows. Boldface lowercase letters and uppercase letters represent vectors and matrices, respectively. Lightface letters represent elements. $\mathbf{A}_{i\cdot}$ and $\mathbf{A}_{\cdot j}$ represent the $i$-th row and the $j$-th column of the matrix $\mathbf{A}$, respectively. $\mathbf{A}_{ij}$ represents the element in the $i$-th row and $j$-th column of $\mathbf{A}$. $(\cdot)^T$ and $(\cdot)^{-1}$ denote the transposition operation and the inverse operation, respectively. $\|\cdot\|_2$ denotes the $\ell_2$ norm.

\section{Background}
In this section, we introduce the system model of uplink asynchronous grant-free CBRA in massive MIMO systems, and SBL for the multi-response case.

\subsection{System Model}
We consider an uplink asynchronous grant-free CBRA in the massive MIMO OFDM system, with a BS located at the centre and $N(N>10^6)$ single-antenna devices located uniformly within a circular area with radius $R$. The BS is equipped with a uniform linear array (ULA) of $M$ antennas and each antenna is separated by half-wavelength. In the same coherent time block, only $K(K\ll N)$ devices are active. Let $a_n\in\{1,0\}$ indicate the user $k$ is active or not. Let's start with the initial grant-free CBRA system model. Each user randomly selects a pilot from the common pilot pool, which is preassigned to the BS. The number and the length of the pilot are $N_p$ and $L$, respectively. The pilot matrix $\mathbf{S}=[\mathbf{s}_1,...\mathbf{s}_{N_p}] \in \mathbb{C}^{L\times N_p}$ consists of these pilots. We assume that the channel is block fading, so it remains constant within a frame. In an ideal scenario where all devices are synchronized, the received signal $\mathbf{U}$ can be expressed as:
\begin{equation}
    \mathbf{U=SD+Z},
    \label{eq1}
\end{equation}
where $\mathbf{U}=[\mathbf{u}_1,\ldots,\mathbf{u}_M]\in\mathbb{C}^{L\times M}$, $\mathbf{D}=[\mathbf{d}_1,\ldots,\mathbf{d}_{N_p}]\in\mathbb{C}^{N_p\times M}$denotes the effective CSI matrix and $\mathbf{Z}\in\mathbb{C}^{L\times M}$ denotes the additive white Gaussian noise
(AWGN) with variance $\sigma_\mathbf{z}^2$ normalized by the transmit power. $\mathbf{u_j}=[u_{1,j},\ldots,u_{L,j}]^T\in\mathbb{C}^{L\times 1}$ is the received signal at the $j$th antenna and $\mathbf{d_i}=[d_{i,1},...,d_{i,M}]\in\mathbb{C}^{1\times M}$ with $d_{i,j}=\alpha_{i,j}\sqrt{g_{i,j}}$ is the $i$-th channel vector that represents the channel coefficient between the $j$-th antenna of the BS and the user who select the $i$-th pilot. $\alpha_{i,j}$ is the normalized Rayleigh fading with zero mean and $g_{i,j}$ is the large-scale fading component. It can be regarded as an MMV sparse signal recovery problem in CS, where $\mathbf{D}$ in (\ref{eq1}) has a row sparsity structure.
\begin{figure}[!t]
\centering
\includegraphics[width=3.5in]{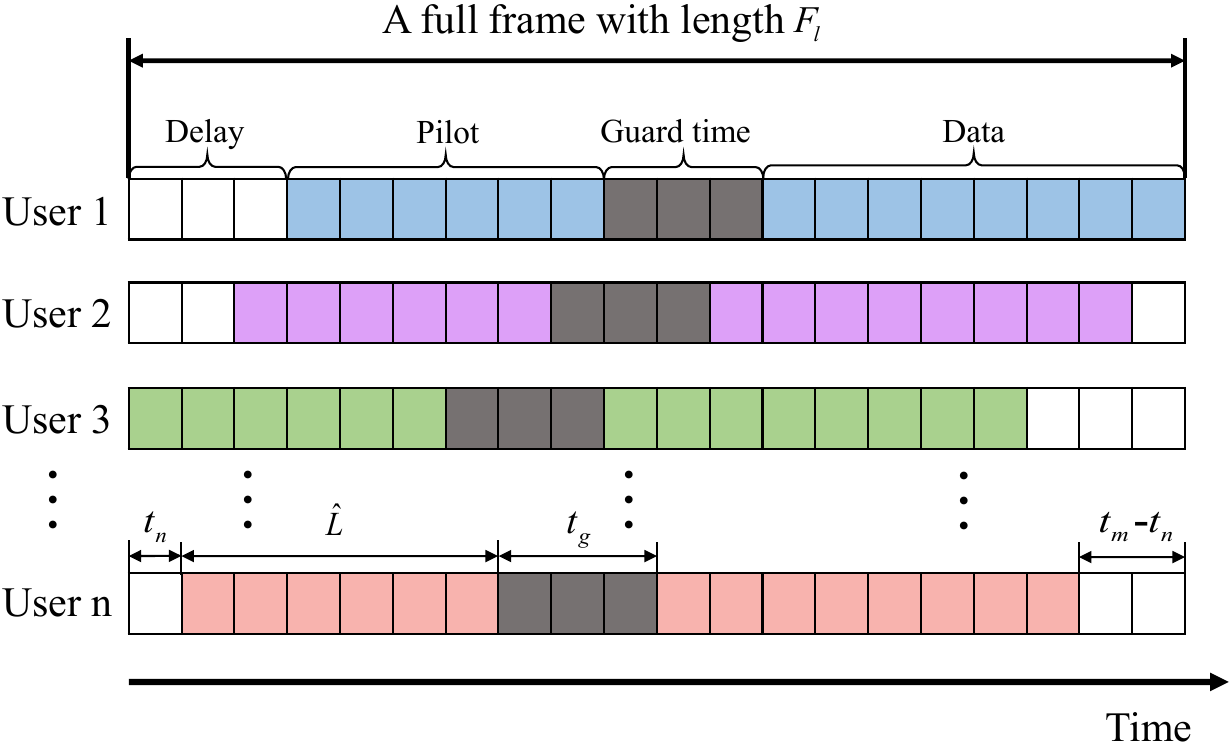}
\caption{The frame structure in asynchronous massive access.}
\label{fig_1}
\end{figure}
\par
For asynchronous situations, the above model is not suitable and some modifications to the model are required. Following the assumption in \cite{bai2023deep,zhu2021deep,chen2022asynchronous}, we consider active users synchronized at the symbol level but asynchronous at the frame level, which is illustrated in Fig. \ref{fig_1}. Specifically, a frame can be divided into multiple symbol intervals, and each symbol is transmitted on the symbol interval. This is reasonable because the symbol interval is the smallest unit in the OFDM system. Therefore, the data of each active user $k$ has an integer symbol delay $t_k$ with $0\leq t_k\leq t_m$, where $t_m$ is the maximum delay of active users. We assume that the maximum symbol delay of active users is much less than the length of the frame $F_l$ and less than the guard time interval $t_{g}$, i.e., $t_m \ll F_l$ and $t_m<t_g$. These features need to be incorporated into the system model. We set $\hat{L}=L+t_m$ and $\hat{N}_p=N_p(t_m+1)$ to simplify the notation, where $\hat{L}$ is the expanded pilot length and $\hat{N}_p$ is the number of expanded pilots in the asynchronous scenario. Next, we expand the $i$-th pilot sequence vector $\mathbf{s}_i$ to the matrix $\hat{\mathbf{S}}_i$ to meet for asynchronous systems, i.e., $\hat{\mathbf{S}}_i=[\hat{\mathbf{s}}_i^0,\ldots,\hat{\mathbf{s}}_i^{t_m}]\in\mathbb{C}^{{\hat{L}}\times (t_m+1)}$, where $\hat{\mathbf{s}}_i^t=[\mathbf{0}^T_{t},\mathbf{s}^T_i,\mathbf{0}^T_{t_m-t}]^T\in\mathbb{C}^{\hat{L}\times 1}$ is the extended pilot sequence that consist of $t$ zeros before $\mathbf{s}_i$ and $(t_m-t)$ zeros after $\mathbf{s}_i$. Then the received signal in the asynchronous scenario can be reformulated as:
\begin{equation}
    \mathbf{\hat{U}=\hat{S}\hat{D}+\hat{Z}},
    \label{eq2}
\end{equation}
where $\hat{\mathbf{U}}=[\hat{\mathbf{u}}_1,...,\hat{\mathbf{u}}_M]\in\mathbb{C}^{\hat{L}\times M}$, $\hat{\mathbf{S}}=[\hat{\mathbf{S}}_1,...,\hat{\mathbf{S}}_{N_p}]$, $\hat{\mathbf{D}}=[\hat{\mathbf{d}}_1;...;\hat{\mathbf{d}}_{\hat{N_p}}]\in\mathbb{C}^{\hat{N}_p\times M}$denotes the effective asynchronous channel matrix and $\hat{\mathbf{Z}}\in\mathbb{C}^{\hat{L}\times M}$ denotes the additive white Gaussian noise
(AWGN) with variance $\hat{\sigma}_\mathbf{z}^2$ normalized by the transmit power. Compared with the synchronous scenario, the joint AD and CE in the asynchronous scenario is more challenging due to the increased dimension of the unknown signal. The asynchronous delay of users is usually treated as a negative factor in existing works on asynchronous CBRA. In contrast, we regard it as an extra degree of freedom in the signal space to enhance massive random access for asynchronous systems. Specifically, users choosing the same pilot could be differentiated by their distinct delays, and thus pilot contention can be resolved in this case.
\par
In addition, since the BS is equipped with multiple antennas and the user is surrounded by rich scatterers, the propagation path of the user has an angular expansion. The channel sparsity in the angular domain can also be exploited to resolve conflicts. Specifically, even if the user chooses the same pilot, as long as their AOAs are different, the cluster-sparse position in the angular domain channel will be different, and the BS can still successfully identify them. The delay-angle domain CSI matrix $\mathbf{\hat{X}}$ can be written as:
\begin{equation}
    \mathbf{\hat{X}}=\mathbf{\hat{D}}\mathbf{\Psi},
    \label{eq3}
\end{equation}
where $\mathbf{\Psi}\in\mathbb{C}^{M\times M}$ is the transform matrix, which is related to the geometry of the array. There are many forms to represent $\mathbf{\Psi}$. In this paper, we take the fast Fourier transformation that is widely used in ULA as a concrete implementation of $\mathbf{\Psi}$. The received signal $\hat{\mathbf{U}}$ can be further expressed in the delay-angle domain as:
\begin{equation}
    \label{eq5}
    \hat{\mathbf{Y}}=\hat{\mathbf{S}}\hat{\mathbf{X}}+\hat{\mathbf{N}},
\end{equation}
where $\mathbf{\hat{Y}=\hat{U}\Psi}=[\mathbf{\hat{y}}_1;\ldots;\mathbf{\hat{y}}_{N_p}]\in\mathbb{C}^{\hat{L}\times M}$, $\mathbf{\hat{X}}=\mathbf{\hat{D}}\mathbf{\hat{\Psi}}=[\mathbf{\hat{x}}_1;\ldots;\mathbf{\hat{x}}_{N_p}]\in\mathbb{C}^{N_p\times M}$ and $\mathbf{\hat{N}=\hat{Z}\Psi}\in\mathbb{C}^{\hat{L}\times M}$.
\par
\begin{figure}[!t]
\centering
\includegraphics[width=3.5in]{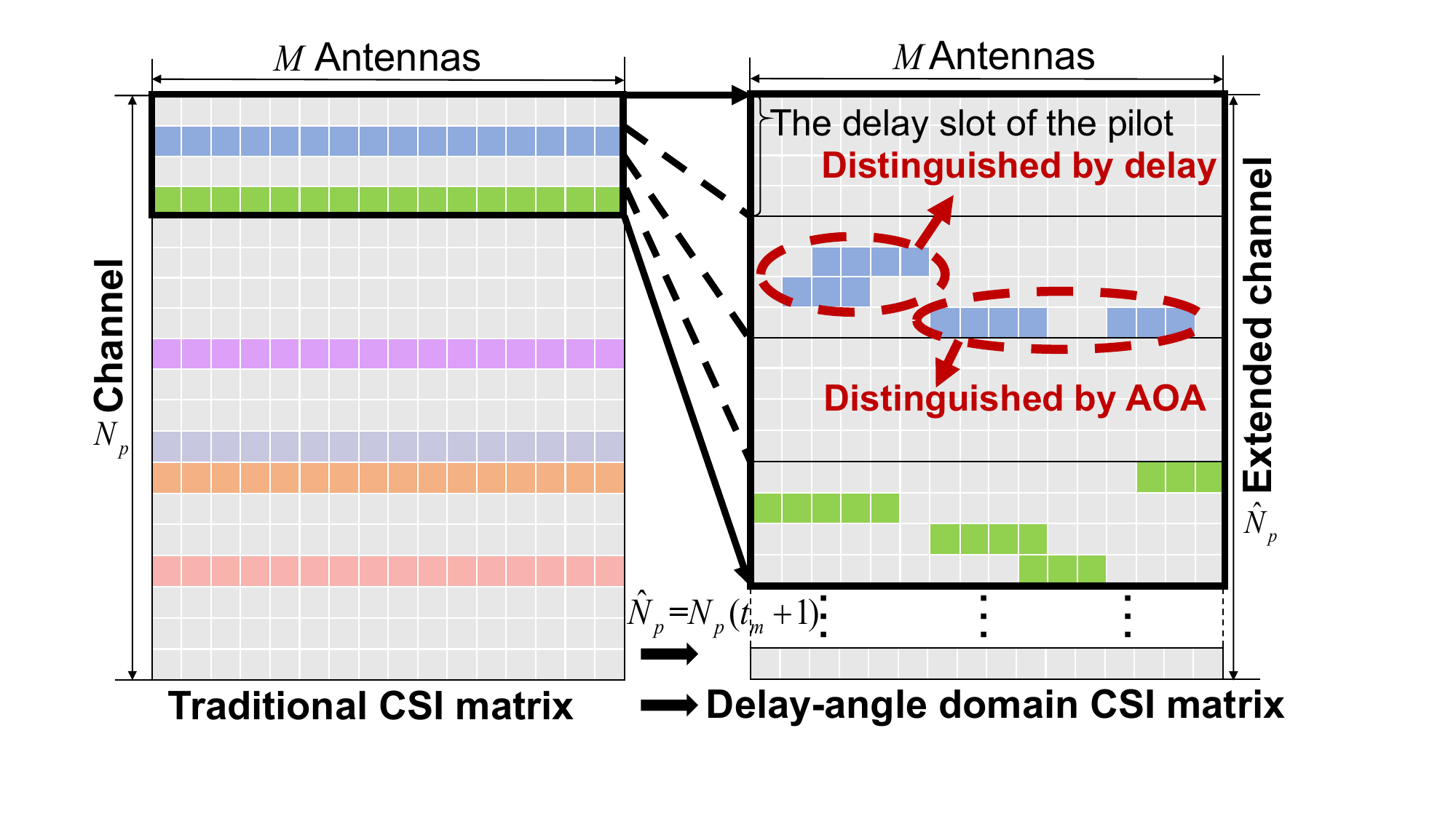}
\caption{An illustration of the traditional MMV model and the proposed delay-angle aided model for joint AD and CE in massive MIMO systems.}
\label{fig_2}
\end{figure}
The delay-angle domain CSI matrix $\hat{\mathbf{D}}$ in the asynchronous scenario and its comparison with the traditional MMV CSI matrix are illustrated in Fig. \ref{fig_2}. It can be seen from the figure that the delay-angle domain CSI matrix has the following structural characteristics: 1) Each row in the CSI matrix $\mathbf{\hat{X}}$ corresponds to an expanded pilot sequence, selected by active users with delay information, resulting in a row-sparse structure for $\mathbf{\hat{X}}$. 2) The number of rows in $\mathbf{\hat{X}}$ increases by a factor of $(t_m+1)$ compared to the original, making $\mathbf{\hat{X}}$ highly row sparse. 3) Since each active user has a distinct angle of arrival (AOA), the non-zero elements in each row of $\mathbf{\hat{X}}$ are clustered within the delay-angle domain, resulting in a highly cluster-sparse structure for the CSI matrix. It is difficult for traditional algorithms to jointly perform AD and CE due to their inability to fully utilize the unique row- and cluster-sparse structures of $\mathbf{\hat{X}}$ in the delay-angle domain, thus presenting a challenge for algorithm design. The algorithm we proposed in the following section is designed specifically to address these structural requirements, optimizing performance in the asynchronous delay-angle domain scenario.
\par
Once our algorithm successfully recovers the CSI matrix $\mathbf{\hat{X}}$, we apply the inverse transformation of (\ref{eq3}) to reconstruct the original CSI matrix $\mathbf{\hat{D}}$, enabling accurate recovery of CSI matrix. The BS can resolve conflicts when users select the same pilot sequence, provided they differ in delay or angle of arrival (AOA). If users share the same pilot, delay, and AOA, they remain undetected. However, the probability of this occurrence is negligible. Our algorithm leverages the additional degrees of freedom provided by the delay-angle domain to resolve conflicts effectively and increase the number of active users supported. Compared to existing grant-free CBRA methods, our approach significantly reduces the additional signaling overhead required for conflict resolution, thereby enhancing pilot efficiency.

\subsection{SBL for Multi-Response Model}
With the base station (BS) equipped with multiple antennas, the problem can be framed as an MMV sparse recovery problem within the CS paradigm. Our algorithm is based on the SBL framework \cite{wipf2004sparse, wipf2007empirical}, which is widely recognized for its effectiveness in sparse recovery problem. Wipf and Rao propose an M-SBL algorithm that allows the maximum sparse subset of design variables to be learned in complex-valued multi-response models. It is demonstrated in\cite{wipf2007empirical} that M-SBL outperforms other multi-response extension methods based on prior Bayesian methods. Considering an MMV sparse recovery problem, the multi-response model of the system in\cite{wipf2007empirical} is:
\begin{equation}
    \mathbf{T=AW+E},
\end{equation}
where $\mathbf{T}=[\mathbf{t}_1,\ldots,\mathbf{t}_M]\in\mathbb{C}^{L\times M}$, $\mathbf{A}\in\mathbb{C}^{L\times N}$, $\mathbf{W}=[\mathbf{w}_1,\ldots,\mathbf{w}_M]\in\mathbb{C}^{N\times M}$ and $\mathbf{E}\in\mathbb{C}^{L\times M}$ denote the system responses, dictionary, unknown sparse matrix and the AWGN noise with variance $\sigma^2$ respectively. For each column of $\mathbf{T}$ and $\mathbf{W}$, the likelihood is
\begin{equation}
\begin{aligned}
    p(\mathbf{t}_j|\mathbf{w}_j)&=(2\pi\sigma^2)^{-M/2}\exp\left(-\frac{1}{2\sigma^2}\|\mathbf{t}_j-\mathbf{A}\mathbf{w}_j\|_2^2\right)\\
    &=\mathcal{N}(\mathbf{A}\mathbf{w}_j,\sigma^2\mathbf{I}_N).
\end{aligned}
\end{equation}
Suppose that the Gaussian prior of $\mathbf{W}$ is
\begin{equation}
    p(\mathbf{W};\boldsymbol{\gamma})\triangleq \mathcal{N}(0,\boldsymbol{\gamma}),
\end{equation}
where $\boldsymbol{\gamma} = [\gamma_1,\ldots,\gamma_N]^T$ is a hyperparameter vector. Each row of $\mathbf{W}$ shares the same hyperparameter $\gamma_i$ to promote row sparsity. Therefore, the posterior distribution of $\mathbf{W}$ is
\begin{equation}
\begin{aligned}
        p(\mathbf{W}|\mathbf{T};\boldsymbol{\gamma})&\propto p(\mathbf{T}|\mathbf{W})P(\mathbf{W};\boldsymbol{\gamma})\\
        &=\mathcal{N}(\boldsymbol{\mathcal{M}},\boldsymbol{\Omega}),
\end{aligned}
\end{equation}
with mean $\boldsymbol{\mathcal{M}}$ and covariance $\boldsymbol{\Omega}$ are
\begin{equation}
  \begin{aligned}
    &\boldsymbol{\mathcal{M}}=\sigma^{-2}\boldsymbol{\Omega}\mathbf{A}\mathbf{Y}\\
    &\boldsymbol{\Omega}=(\sigma^{-2}\mathbf{A}^T\mathbf{A}+\boldsymbol{\Gamma})^{-1},
  \end{aligned}
\end{equation}
respectively, where $\boldsymbol{\Gamma}\triangleq\mathrm{diag}(\boldsymbol{\gamma})$. The maximum a posterior (MAP) estimate of $\mathbf{W}$ is the mean of its posterior distribution:
\begin{equation}
    \label{eq4}
    \mathbf{X}_{MAP}=(\mathbf{A}^T\mathbf{A}+\sigma^{-2}\boldsymbol{\Gamma})^{-1}\mathbf{A}\mathbf{Y}.
\end{equation}
Therefore, the problem becomes to estimate the hyperparameter vector $\boldsymbol{\gamma}$ in (\ref{eq4}). The EM algorithm\cite{dempster1977maximum,moon1996expectation} is used for hyperparameter estimation due to its excellent performance. According to the EM algorithm, the update rule of $\boldsymbol{\gamma}$ is:
\begin{equation}
    \gamma_i^{(t)}=\frac{1}{M}\|\boldsymbol{\mathcal{M}}_{i\cdot}\|_2^2+\boldsymbol{\Omega}_{ii},\quad\forall i=1,\ldots,N.
\end{equation}
In addition to the SBL-based approach, other advanced methods are available for sparse recovery in multi-measurement scenarios, such as M-BP \cite{malioutov2005sparse}, M-FOCUSS \cite{cotter2005sparse}, and the $\ell_1/\ell_2$ reweighting algorithm \cite{chen2021joint}. In this paper, we extend the expectation-maximization (EM) algorithm to solve for hyperparameters while incorporating cluster sparsity. This EM-based extension is specifically tailored to handle the cluster-sparse structure inherent in the delay-angle domain, setting it apart from other sparse recovery methods.

\section{Cluster-Extended SBL for Asynchronous Grant-Free CBRA}
In this section, we formulate our problem based on the system model (\ref{eq5}) in the asynchronous grant-free CBRA scenario. We develop a CE-SBL algorithm that simultaneously utilizes row and cluster sparsity for AD and CE. We will explain the derivation of the algorithm in detail. 
\subsection{Problem Formulation}
To effectively handle cluster-sparse signals with an unknown block-sparse structure, we introduce a joint sparse prior that leverages both row and cluster-sparse characteristics. This weighted prior combines row-sparse and cluster-sparse priors, statistically correlating with the sparse patterns of adjacent coefficients. As a result, each element of $\mathbf{\hat{X}}$ has an individual prior influenced by the coefficients in adjacent rows. Thus, the prior of $\mathbf{\hat{X}}$ can be expressed as:
\begin{equation}
    p(\mathbf{\hat{X}};\boldsymbol{\Gamma})=p(\mathbf{\hat{X}};\boldsymbol{\alpha},\boldsymbol{\beta})=\prod_{i=1}^{\hat{N}}\prod_{J=1}^{M}\mathcal{N}(0,\gamma_{i,j}^{-1}),
    \label{eq6}
\end{equation}
with $\gamma_{i,j}^{-1}=\lambda\alpha_i^{-1}+(1-\lambda)(\kappa\beta_{i,j-1}+\beta_{i,j}+\kappa\beta_{i,j+1})^{-1}$, where $\boldsymbol{\alpha}$ and $\boldsymbol{\beta}$ are hyperparameters that capture the row-sparse and cluster-sparse characteristics of $\mathbf{\hat{X}}$ respectively. $\lambda\in[0,1]$ is a trade-off parameter used to modulates the influence of $\alpha$ and $\beta$. $\kappa>0$ is also a parameter that indicates the correlation of $\beta_{i,j}$ with its adjacent hyperparameters $\beta_{i,j-1}$ and $\beta_{i,j+1}$. The elements of $i$th row share the same $\alpha_i$. To simplify the notation, we set $\tilde\beta_{i,j}=\kappa\beta_{i,j-1}+\beta_{i,j}+\kappa\beta_{i,j+1}$. When $\lambda=1$, the prior for $\mathbf{\hat{X}}$ simplifies to that of the M-SBL algorithm as described in \cite{wipf2007empirical}, indicating a purely row-sparse structure. Conversely, when $\lambda=0$, the prior aligns closely with the PCSBL approach detailed in \cite{fang2014pattern}, emphasizing a cluster sparsity. Since the physical AOA is constrained within the range $[-\pi/2,\pi/2]$, we define the periodic boundary conditions as $\gamma_{i,0}=\gamma_{i,M}$ and $\gamma_{i,M+1}=\gamma_{i,1}$ in (\ref{eq6}). It is worth mentioning that the PCSBL utilizes the column prior to solving the single measurement vector (SMV) problem. When considering MMV problems, it is generally adopted to transform the MMV problem into the SMV problem by Kronecker product\cite{fang2014pattern2}, which will significantly improve the algorithm complexity when solving large-scale problems such as massive random access, especially in the matrix inversion operation. A primary advantage of our prior design is the direct use of row-cluster priors in our algorithm, which reduces computational complexity. But the problem becomes more difficult to solve than before. i.e., we trade more difficult algorithm designs for less complexity. Specifically, our algorithm needs to combine such row-cluster priors and the original row priors to deal with such prior. Another advantage of using this prior in (\ref{eq6}) is that the hyperparameters are coupled together by trade-off coefficients $\lambda$ and correlation coefficients $\kappa$, thus the cluster character of the signal can be learned automatically without any additional prior information about the structure of the signal.
\par
Assumed that the hyperprior of the parameters follows the Gamma distribution
\begin{equation}
\begin{aligned}
    &p(\boldsymbol\alpha)=\prod\limits_{i=1}^{\hat{N}}\mathrm{Gamma}(\alpha_i|a,b)=\prod\limits_{i=1}^{\hat{N}}\Gamma(a)^{-1}b^a\alpha_i^ae^{-b\alpha_i}\\
    &p(\boldsymbol\beta)=\prod\limits_{i=1}^{\hat{N}}\prod\limits_{i=1}^M\mathrm{Gamma}(\beta_{i,j}|c,d)=\prod\limits_{i=1}^{\hat{N}}\prod\limits_{i=1}^M\Gamma(c)^{-1}d^c\beta_{i,j}^ce^{-d\beta_{i,j}},
\end{aligned}
    \label{eq8}
\end{equation}
and the noise is AWGN with variance $\sigma_{\hat{z}}^2$, where $\Gamma(x)=\int_{0}^{\infty}t^{x-1}e^{-t}dt$ is the Gamma function. In classical SBL theory, it is common to give very small values to the hyperparameter set $\{a,b,c,d\}$, e.g., $10^{-4}$. Similarly, we give the same values to parameters $b$ and $d$. However, the difference is that we will give a larger value to $a$ and $c$. Because a larger parameter value will make the corresponding variance smaller, thus promoting the sparsity of the solution. The specific values chosen for $a$ and $c$, along with their impact on performance, will be analyzed in detail in Section IV. The likelihood of each column of $\mathbf{\hat{Y}}$ is given by
\begin{equation}
\begin{aligned}
    p(\mathbf{\hat{y}}_j|\mathbf{\hat{x}}_j)&=(2\pi\sigma^2)^{-M/2}\exp\left(-\frac{1}{2\sigma^2}\|\mathbf{\hat{y}}_j-\mathbf{A}\mathbf{\hat{x}}_j\|_2^2\right)\\
    &=\mathcal{N}(\mathbf{A}\mathbf{\hat{x}}_j,\sigma^2\mathbf{I}_N).
\end{aligned}
\end{equation}
Combined with $p(\mathbf{\hat{x}}_j|\mathbf{\hat{y}}_ j;\boldsymbol{\gamma}_j)\propto p(\mathbf{\hat{x}};\boldsymbol{\gamma}_j)p(\mathbf{\hat{y}}_j|\mathbf{\hat{x}}_j)$, we compute the posterior distribution of $j$th column of $\mathbf{\hat{X}}$ which enables us to update the estimated sparse coefficients iteratively based on observed data:
\begin{equation}
    p(\mathbf{\hat{x}}_j|\mathbf{\hat{y}}_j;\boldsymbol{\gamma}_j)=\mathcal{N}(\boldsymbol{\mu}_j,\boldsymbol{\Omega}_{j})
\end{equation}
with 
\begin{equation}
\begin{aligned}
    &\boldsymbol{\mu}_j=\sigma_{\hat{z}}^{-2}\boldsymbol{\Omega}_j\mathbf{A}\mathbf{\hat{y}}_j,\\
    &\boldsymbol{\Omega_j}=(\sigma^{-2}\mathbf{A}^T\mathbf{A}+\boldsymbol{\tilde{\Gamma}}_j)^{-1},
\end{aligned}
\label{eq10}
\end{equation}
where $\boldsymbol{\mu}_j\in\mathbb{C}^{\hat{N}\times 1}$, $\boldsymbol{\Omega_j}\in\mathbb{C}^{\hat{N}\times \hat{N}}$, and $\boldsymbol{\tilde{\Gamma}}_j=\mathrm{diag}(\boldsymbol{\gamma_j})$. We set $\boldsymbol{\phi}_j=\mathrm{diag}(\boldsymbol{\Omega_j})$ By combining $M$ posterior distributions, we get $\boldsymbol{\mathcal{M}}=[\boldsymbol{\mu}_1,\ldots,\boldsymbol{\mu}_M]\in\mathbb{C}^{\hat{N}\times M}$. Thus, the MAP estimate is the mean of the posterior distribution of $\mathbf{\hat{X}}$, which is
\begin{equation}
    \mathbf{\hat{X}}_{MAP}=\boldsymbol{\mathcal{M}}.
    \label{eq7}
\end{equation}
Our problem now is to estimate the hyperparameter $\boldsymbol{\gamma}$ in (\ref{eq7}) to determine the optimal value of the MAP estimation.

\subsection{Cluster-Extended SBL Algorithm Derivation}
To accurately joint AD and CE in asynchronous CBRA systems, we need to solve the MAP estimation problem in (\ref{eq7}). We can see that the problem is non-convex and there is no closed-form solution, necessitating an iterative approach. In order to solve this problem more effectively, we propose a CE-SBL algorithm. We apply the paradigm of the EM algorithm to iteratively solve the lower bound of maximizing the posterior probability, where parameters other than hyperparameters are treated as hidden variables. Note that $\boldsymbol{\alpha}$ and $\boldsymbol{\beta}$ in hyperparameter $\boldsymbol{\gamma}$ are coupled, which aggravates the analysis. The EM algorithm alternates between E-steps and M-steps. The E-step is to fix the hyperparameter and calculates the expected value of the latent variable and the M-step fixes the latent variables and optimizes the hyperparameters to maximize the expectations derived in the E-step. The derivation of the CE-SBL algorithm to solve this problem is detailed as follows::
\subsubsection{E-step}
In the E-step, we calculate the posterior expected value of the latent variable. Specifically, we employ the Q-function as the expected value of the objective function. The Q-function represents the expected value of the conditional probability distribution of the latent variable $\mathbf{\hat{X}}$ given the observed data $\mathbf{\hat{Y}}$ and the current parameter estimates $\boldsymbol{\Gamma}^t$: 
\begin{equation}
\begin{aligned}
    Q(\boldsymbol{\Gamma},\boldsymbol{\Gamma}^{t})&=\mathbb{E}_{\mathbf{\hat{X}}|\mathbf{\hat{Y}},\Gamma^t}[\log p(\Gamma|\mathbf{\hat{Y}},\mathbf{\hat{X}})]\\
    &\propto\int p(\mathbf{\hat{X}}|\mathbf{\hat{Y}},\Gamma^t)\log[p(\boldsymbol{\Gamma})p(\mathbf{\hat{X}}|\boldsymbol{\Gamma})]d\mathbf{\hat{X}}.
\end{aligned}
\end{equation}
Extracting terms independent of $\mathbf{\hat{X}}$, we obtain:
\begin{equation}
\begin{aligned}
    Q(\boldsymbol{\Gamma},\boldsymbol{\Gamma}^{t})&=\log p(\boldsymbol{\Gamma})+\int p(\mathbf{\hat{X}}|\mathbf{\hat{Y}},\Gamma^t)\log p(\mathbf{\hat{X}}|\boldsymbol{\Gamma})d\mathbf{\hat{X}}\\
    &=\log (p(\boldsymbol{\alpha})p(\boldsymbol{\beta}))+\int p(\mathbf{\hat{X}}|\mathbf{\hat{Y}},\Gamma^t)\log p(\mathbf{\hat{X}}|\boldsymbol{\Gamma})d\mathbf{\hat{X}}.
\end{aligned}
\label{eq9}
\end{equation}
By substituting (\ref{eq6}) and (\ref{eq8}) into (\ref{eq9}), the Q-function is expressed as:
\begin{equation}
\begin{aligned}
    &Q(\boldsymbol{\Gamma},\boldsymbol{\Gamma}^{t})\\=
    &\sum_{i=1}^{\hat{N}}\sum_{j=1}^{\hat{M}}\Bigg(a\log\alpha_{i}+b\beta_{i,j}-c\alpha_{i}-d\beta_{i,j}\\
    &\quad+\frac{1}{2}\log\Big(\lambda\alpha_{i}^{-1}+(1-\lambda)\beta_{i,j}^{-1}\Big)^{-1}\\
    &\quad-\frac{1}{2}\Big(\lambda\alpha_{i}^{-1}+(1-\lambda)\beta_{i,j}^{-1}\Big)^{-1}\int p(\mathbf{\hat{X}}|\mathbf{\hat{Y}},\boldsymbol{\Gamma}^{t})\hat{x}_{i,j}^2d\mathbf{\hat{X}}\Bigg).
\end{aligned}
\label{eq11}
\end{equation}
where $\hat{x}_{i,j}$ denotes the $i$-th element of $\mathbf{\hat{x}}_{j}$. Given the derivation of the mean $\boldsymbol{\mathcal{M}}$ and variance $\boldsymbol{\tilde{\Omega}}$ of the posterior distribution of $\mathbf{\hat{X}}$ in (\ref{eq10}), we obtain:
\begin{equation}
\begin{aligned}
    \int p(\mathbf{\hat{X}}|\mathbf{\hat{Y}},\boldsymbol{\Gamma}^{t})\hat{x}_{i,j}^2d\mathbf{\hat{X}}&=\mathbb{E}_{\mathbf{\hat{X}}|\mathbf{\hat{Y}},\boldsymbol{\Gamma}^t}[\hat{x}_{i,j}^2]\\
    &=\hat{\mu}_{i,j}^2+\hat{\phi}_{i,j},
\end{aligned}
\label{eq12}
\end{equation}
where $\mu_{i,j}$ and $\phi_{i,j}$ denote the $i$-th element of $\boldsymbol{\mu}_j$ and $\boldsymbol{\phi}_j$, respectively. By substituting (\ref{eq12}) into (\ref{eq11}), the final form of the Q-function is obtained as:
\begin{equation}
\begin{aligned}
    Q(\boldsymbol{\Gamma},\boldsymbol{\Gamma}^{t})&=\sum_{i=1}^{\hat{N}}\sum_{j=1}^{\hat{M}}\Bigg(a\log\alpha_{i}+c\log\beta_{i,j}-b\alpha_{i}-d\beta_{i,j}\\
    &\quad+\frac{1}{2}\log\Big(\lambda\alpha_{i}^{-1}+(1-\lambda)\beta_{i,j}^{-1}\Big)^{-1}\\
    &\quad-\frac{1}{2}\Big(\lambda\alpha_{i}^{-1}+(1-\lambda)\beta_{i,j}^{-1}\Big)^{-1}(\mu_{i,j}^2+\phi_{i,j})\Bigg).
\end{aligned}
    \label{eq13}
\end{equation}

\subsubsection{M-step}
The objective of the M-step is to update the hyperparameters $\boldsymbol{\Gamma}$ to maximize the Q-function, which is defined as:
\begin{equation}
    \boldsymbol{\Gamma}^{t+1}=\arg\max_{\boldsymbol{\Gamma}}Q(\boldsymbol{\Gamma},\boldsymbol{\Gamma}^{t}).
\end{equation}
The section outlines the strategy for updating the hyperparameters.
\par
The hyperparameters $\boldsymbol{\alpha}$ and $\boldsymbol{\beta}$ are coupled by a logarithmic term, presenting a difficult challenge. As the Q-function does not have a closed-form solution, numerical methods need to be employed. While an iterative algorithm could be employed to optimize the Q-function, it would significantly increase computational time and reduce efficiency. Because the M-step of each EM algorithm iteration also requires an iterative approach to solve the Q-function, this results in a nested iterative process. The actual experiment also confirmed our idea. Therefore, to improve computational efficiency, we apply appropriate approximations to decouple the hyperparameters, allowing them to be solved separately. Although approximate strategies may introduce some error, careful application ensures that the error remains within acceptable tolerance for the problem. Specifically, we apply Jensen’s inequality to approximate the reciprocal and logarithmic terms in (\ref{eq13}), disregarding the Jensen gap. This yields:
\begin{equation}
    \Big(\lambda\alpha_{i}^{-1}+(1-\lambda)\beta_{i,j}^{-1}\Big)^{-1}\approx\frac{\alpha_{i}}{4\lambda}+\frac{\beta_{i,j}}{4(1-\lambda)},
\end{equation}
\begin{equation}
    \log\Big(\lambda\alpha_{i}^{-1}+(1-\lambda)\beta_{i,j}^{-1}\Big)^{-1}\approx\frac{1}{2}\log\frac{\alpha_{i}}{4\lambda}+\frac{1}{2}\log\frac{\beta_{i,j}}{4(1-\lambda)}.
\end{equation}
The approximation is reasonable. Since the solution to the problem is sparse and in most positions the prior parameters $\boldsymbol{\alpha}$ and $\boldsymbol{\beta}$ are very large values, e.g., $10^6$. In this case, $\boldsymbol{\alpha}$ and $\boldsymbol{\beta}$ are relatively close, so the arithmetic mean and geometric mean introduces minimal error. As a result, equation (\ref{eq13}) simplifies to:
\begin{equation}
    Q(\boldsymbol{\Gamma},\boldsymbol{\Gamma}^{t})=Q_1(\boldsymbol{\alpha},\boldsymbol{\alpha}^{t})+Q_2(\boldsymbol{\beta},\boldsymbol{\beta}^{t})
\end{equation}
with the two part $Q_1$ and $Q_2$ given respectively by
\begin{equation}
\begin{aligned}
    Q_1(\boldsymbol{\alpha},\boldsymbol{\alpha}^{t})&=\sum_{i=1}^{\hat{N}}\sum_{j=1}^M\Bigg(a\log\alpha_{i}-b\alpha_{i}+\frac{1}{4}\log\frac{\alpha_{i}}{4\lambda}\\
    &\quad-\frac{\alpha_{i}}{8\lambda}(\mu_{i,j}^2+\phi_{i,j})\Bigg),
\end{aligned}
\end{equation}
\begin{equation}
\begin{aligned}
    &Q_2(\boldsymbol{\beta},\boldsymbol{\beta}^{t})\\    
    =&\sum_{i=1}^{\hat{N}}\sum_{j=1}^{M}\Bigg(c\log\beta_{i,j}-d\beta_{i,j}+\frac{1}{4}\log\frac{\beta_{i,j}}{4(1-\lambda)}\\
    &\quad-\frac{1}{8(1-\lambda)}\beta_{i,j}(\mu_{i,j}^2+\phi_{i,j})\Bigg)\\
    =&\sum_{i=1}^{\hat{N}}\sum_{j=1}^{M}\Bigg(\frac{1}{4}\log\Big(\frac{\kappa\beta_{i,j-1}+\beta_{i,j}+\kappa\beta_{i,j+1}}{4(1-\lambda)}\Big)\\
    &\quad-\frac{1}{8(1-\lambda)}(\kappa\beta_{i,j-1}+\beta_{i,j}+\kappa\beta_{i,j+1})(\mu_{i,j}^2+\phi_{i,j})\\
    &\quad+c\log\beta_{i,j}-d\beta_{i,j}\Bigg).
\end{aligned}
\label{eq14}
\end{equation}
Our strategy now is to optimize $Q_1$ and $Q_2$ separately.
\paragraph{Optimization of $Q_1$ function}
Since the hyperparameter $\boldsymbol{\alpha}$ in the $Q_1$ function is independent and the $Q_1$ function is convex, its exact solution can be obtained using the first-order optimality condition. Taking the derivative of $Q_1$ with respect to $\alpha_{i}$, we obtain:
\begin{equation}
    \frac{\partial Q(\boldsymbol{\alpha},\boldsymbol{\alpha}^{t})}{\partial\boldsymbol{\alpha}_{i}}=\left(\frac{a}{\alpha_i}-b+\frac{1}{4\alpha_i}\right)M-\frac{1}{8\lambda}\sum_{j=1}^M(\mu_{i,j}^2+\phi_{i,j}).
\end{equation}
Set the above equation equal to zero, and we end up with:
\begin{equation}
    \alpha_i=\frac{8a+2}{8b+v_i/\lambda},\quad\forall i=1,\ldots,\hat{N}.
\label{eq20}
\end{equation}
where $v_i=\frac{1}{M}\sum_{j=1}^M(\mu_{i,j}^2+\phi_{i,j})$.

\paragraph{Optimization of $Q_2$ function}
For the optimization of the Q2 function, The challenge arises from the coupling of each hyperparameter $\beta_{i,j}$ with its adjacent hyperparameters $\beta_{i,j-1}$ and $\beta_{i,j+1}$ along the row, due to the logarithmic term $\log\Big((1-\lambda)(\kappa\beta_{i,j-1} + \beta_{i,j} + \kappa\beta_{i,j+1})\Big)$. If we use an iterative algorithm, such as gradient descent, it would require a nested iterative process, resulting in high computational complexity. To address this, we employ an alternative strategy to obtain the solution for the $Q_2$ function. Specifically, we leverage the first-order optimality condition along with an appropriate approximation strategy to obtain the solution.
\par
To simplify the notation, we rewrite (\ref{eq14}) in vector form and omit the constant term:
\begin{equation}
\begin{aligned}
    &Q_2(\boldsymbol{\beta},\boldsymbol{\beta}^{t})\\
    =&\sum_{j=1}^{M}\Bigg(\frac{1}{4}\log(\kappa\boldsymbol{\beta}_{j-1}+\boldsymbol{\beta}_{j}+\kappa\boldsymbol{\beta}_{j+1})\\
    &\quad-\frac{1}{8(1-\lambda)}(\kappa\boldsymbol{\beta}_{j-1}+\boldsymbol{\beta}_{j}+\kappa\boldsymbol{\beta}_{j+1})(\boldsymbol{\mu}_{\cdot j}^2+\boldsymbol{\phi}_{j})\\
    &\quad+c\log\boldsymbol{\beta}_{j}-d\boldsymbol{\beta}_{j}\Bigg).
\end{aligned}
\label{eq15}
\end{equation}
Take the first derivative of (\ref{eq15}) with respect to $\boldsymbol{\beta_j}$, we have
\begin{equation}
    Q(\boldsymbol\beta|\boldsymbol\beta^{t})^{\prime}=\frac{1}{4}(\kappa\mathbf{p}_{j-1}+\mathbf{p}_j+\kappa\mathbf{p}_{j+1})-\frac{\mathbf{q}_j}{8(1-\lambda)}+\frac{c}{\boldsymbol{\beta}_j}-d
\end{equation}
with
\begin{equation}
    \mathbf{p}_j=\frac{1}{\kappa\boldsymbol{\beta}_{j-1}+\boldsymbol{\beta}_j+\kappa\boldsymbol{\beta}_{j+1}},
    \label{eq16}
\end{equation}
\begin{equation}
    \mathbf{q}_j=\kappa\boldsymbol{\mu}_{j-1}^2+\boldsymbol{\mu}_j^2+\kappa\boldsymbol{\mu}_{j+1}^2+\kappa\boldsymbol{\phi_{j-1}}+\boldsymbol{\phi_j}+\kappa\boldsymbol{\phi_{j+1}}.
\end{equation}
where $\boldsymbol{\beta}_{-1}=\boldsymbol{\beta}_{M-1}$ and $\boldsymbol{\beta}_{M+2}=\boldsymbol{\beta}_{2}$ since the physical AOA is cyclical in the range $[-\pi/2,\pi/2]$. Set the first derivative equal to 0, and we obtain the following equation:
\begin{equation}
    \mathbf{p}_j+\kappa\mathbf{p}_{j+1}+\kappa\mathbf{p}_{j-1}=\frac{\mathbf{q}_j}{2(1-\lambda)}-\frac{4c}{\boldsymbol{\beta}_j}+4d.
    \label{eq19}
\end{equation}
Note that in (\ref{eq16}), the hyperparameter set \{$\boldsymbol{\beta}_{j-1}$, $\boldsymbol{\beta}_j$, $\boldsymbol{\beta}_{j+1}$\} and the correlation coefficient $\kappa$ are positive, thus we have
\begin{equation}
    \mathbf{p}_j<\frac{1}{\kappa\boldsymbol{\beta}_{j-1}},\mathbf{p}_j<\frac{1}{\boldsymbol{\beta}_j},\mathbf{p}_j<\frac{1}{\kappa\boldsymbol{\beta}_{j-1}}.
    \label{eq17}
\end{equation}
Similarly, by changing the index of (\ref{eq17}), we get the following upper bound:
\begin{equation}
    \mathbf{p}_{j-1}<\frac{1}{\kappa\boldsymbol{\beta}_j},\mathbf{p}_j<\frac{1}{\boldsymbol{\beta}_j},\mathbf{p}_{j+1}<\frac{1}{\kappa\boldsymbol{\beta}_j}.
    \label{eq18}
\end{equation}
Substituting (\ref{eq18}) and $\boldsymbol{p}_j>0$ into (\ref{eq19}), we get the upper and lower bound on $\boldsymbol{\beta}_j$, respectively:
\begin{equation}
    \frac{8c}{8d+\mathbf{q}_j/(1-\lambda)}<\boldsymbol{\beta}_j<\frac{8c+6}{8d+\mathbf{q}_j/(1-\lambda)}.\quad\forall j=1,\ldots,M.
\end{equation}
Empirically, we take the lower bound as the parameter estimate $\boldsymbol{\beta}$, i.e.,
\begin{equation}
    \boldsymbol{\beta}_{i,j}=\frac{8c}{8d+\mathbf{q}_{i,j}/(1-\lambda)},\quad\forall i=1,\ldots,\hat{N},j=1,\ldots,M.
    \label{eq21}
\end{equation}
where $\mathbf{q}_{i,j}$ is the $i$-th element of $\mathbf{q}_j$. At this point, we have completed the estimation of all hyperparameters. Combining $\boldsymbol{\alpha}$ and $\boldsymbol{\beta}$ in (\ref{eq20}) and (\ref{eq21}), the final hyperparameter estimate is
\begin{equation}
\begin{aligned}
    \gamma_{i,j}^{-1}=\lambda\alpha_i^{-1}+(1-\lambda)(\kappa\beta_{i,j-1}+\beta_{i,j}+\kappa\beta_{i,j+1})^{-1},\\
    \forall i=1,\ldots,\hat{N},j=1,\ldots,M.
\end{aligned}
\label{eq22}
\end{equation}
\par
Finally, by iterating through the update rules above, we derive the CE-SBL algorithm for joint AD and CE, summarized in Algorithm \ref{algorithm1}. Although a suboptimal strategy is employed to update the hyperparameters in the M-step, subsequent numerical experiments demonstrate the effectiveness of the strategy. It achieves performance comparable to gradient-based methods for the M-step update, while significantly reducing complexity.  This is because when the prior parameters are non-zero, combined with the convergence conditions of the EM algorithm, our suboptimal solution successfully provides a reasonable optimal solution estimation and ensures convergence to the correct solution.
\begin{algorithm}[H]
\caption{CE-SBL}\label{algorithm1}
\begin{algorithmic}
\STATE 
\STATE step 1: Initialize all parameters with $\lambda=0.01$, $\kappa=0.01$,
\STATE \hspace{1.1cm}$\boldsymbol{\Gamma}=\textbf{1}$ and $t_m$;
\STATE step 2: Outer-loop. Compute the mean $\boldsymbol{\mathcal{M}}$ and variance $\boldsymbol{\Omega}$
\STATE \hspace{1.1cm}of the posterior probability by (\ref{eq10}) and the MAP
\STATE \hspace{1.1cm}estimate of $\mathbf{\hat{X}}_{MAP}$ by (\ref{eq7});
\STATE step 3: Inner-loop. Update the hyperparameters $\boldsymbol{\alpha}^{t+1}$ and
\STATE \hspace{1.1cm}$\boldsymbol{\beta}^{t+1}$ using (\ref{eq20}) and (\ref{eq21}) respectively, and combine
\STATE \hspace{1.1cm}them by (\ref{eq22}) to update $\boldsymbol{\Gamma}^{t+1}$;
\STATE step 4: Keep iterating step 2 and step 3 until $\|\hat{\mathbf{X}}^{t+1}-\hat{\mathbf{X}}^{t}\|_2^2$
\STATE \hspace{1.1cm}$<10^{-8}$, and finally get the solution for $\mathbf{\hat{X}}$.
\end{algorithmic}
\end{algorithm}
\subsection{Analysis}
It is well-established that the performance of sparse linear inverse problem (LIP) solutions is closely related to the sparsity of the signal $\hat{\mathbf{X}}$. Specifically, an increase in active users results in reduced recovery accuracy. We demonstrate that the proposed framework can support a greater number of active users compared to traditional frameworks, thereby significantly increasing system capacity.
Before going into specifics, it is necessary to state the following lemma from \cite{cotter2005sparse}, which indicates the maximum sparsity supported in a noiseless MMV problem.
\par
\textit{Lemma 1:} For a noiseless MMV problem, we can write it as $\hat{\mathbf{U}}=\hat{\mathbf{S}}\hat{\mathbf{D}}$, where $\hat{\mathbf{U}}\in\mathbb{C}^{\hat{L}\times M}$, $\hat{\mathbf{S}}\in\mathbb{C}^{\hat{L}\times \hat{N}}$, and $\hat{\mathbf{D}}\in\mathbb{C}^{\hat{N}\times M}$. Any $\hat{L}$ columns of $\hat{\mathbf{S}}$ are linearly independent and $rank(\mathbf{\hat{U}})=\hat{M}\leq\hat{L}$. There exists a unique sparse solution $\hat{\mathbf{D}}$ with sparsity $r$ and $r\leq\lceil(\hat{L}+\hat{M})/2\rceil-1$, where $\lceil\cdot\rceil$ is the ceiling operation.
\par
Our framework transforms the MMV problem in traditional massive random access scenarios to the delay-angle domain. Theorem 1 demonstrates that our proposed framework exhibits greater sparsity than traditional frameworks, allowing it to support a larger number of active users.
\par
\textit{Theorem 1:} For a noiseless sparse LIP with unknown cluster patterns, we can write it as $\hat{\mathbf{Y}}=\hat{\mathbf{S}}\hat{\mathbf{X}}$, where $\hat{\mathbf{Y}}\in\mathbb{C}^{\hat{L}\times M}$, $\hat{\mathbf{S}}\in\mathbb{C}^{\hat{L}\times \hat{N}}$, and $\hat{\mathbf{X}}\in\mathbb{C}^{\hat{N}\times M}$. Any $\hat{L}$ columns of $\hat{\mathbf{S}}$ are linearly independent and $rank(\mathbf{\hat{Y}})=\hat{M}\leq\hat{L}$. Assume that the maximum length of the cluster in $\hat{\mathbf{X}}$ is $D$, where $D\ll M$, then there exists a unique sparse solution $\hat{\mathbf{X}}$ with sparsity $\hat{r}$ and $\hat{r}\leq\lfloor M/D\rfloor(\lceil(\hat{L}+D)/2\rceil-1)$, where $\lfloor\cdot\rfloor$ is the flooring operation and $\lceil\cdot\rceil$ is the ceiling operation. And the maximum value $\hat{r}_m$ of $\hat{r}$ is greater than that in Lemma 1.
\par
\textit{Proof:} See Appendix.
\par
Theorem 1 indicates that a larger maximum sparsity can be achieved through the delay-angle domain. This means that our framework can support a much larger number of active users than traditional MMV frameworks in general massive access scenarios.

\section{Simulation Results}
In this section, we conduct various numerical experiments to illustrate the gains of our proposed framework and algorithm with different settings. Our initial parameters are set as follows unless otherwise specified: We consider $10^6$ potential users distributed across a BS within a coverage radius of $0.5$ kilometers. The BS is equipped with $M=64$ antennas and each device has one antenna. The common pilot pool assigned to the BS contains $N=64$ pilots. There are $K=30$ active users within the BS coverage area. When the user has data to transmit, the active user randomly selects a pilot from the common pilot pool. The pilot length is set to $L=64$. We consider an OFDM system with a carrier frequency of 2GHz, a channel bandwidth of $B=1.4MHz$, and a total of $72$ subcarriers for uplink communication. The SNR of the system is set to $15$dB. We use the one-ring channel model formulated in \cite{zhou2007experimental}, where the channel coefficient of user $k$ is modelled as:
\begin{equation}
    \mathbf{h}_k=\sum\limits_{l=1}^{L_{path}}g_{k,l}\mathbf{a}(\phi_{k,l})e^{j\pi\tau_{k,l}B},
\end{equation}
where $L_{path}=16$ is the spreading paths, $g_{k,l}\sim\mathcal{CN}(g;0,1)$ and $\tau_{k,l}\sim\mathcal{U}(0, L_{cp}/B)$ are the gain and the delay of each spreading path of user $k$, respectively. $\mathcal{CN}(:;0,1)$ denotes the complex Gaussian distribution with mean $0$ and variance $1$, $\mathcal{U}(0,L_{cp}/B)$ denotes the uniform distribution from $0$ to $L_{cp}/B$, and $L_{cp}=64$ is the length of the cyclic prefix. Similar to the 3GPP technical report \cite{3gpp}, we consider the user's AOA in a cluster. The array response vector is given by $\mathbf{a}(\phi_{k,l})=[1,e^{-j2\pi\frac{d}{\lambda}sin\varphi_{k,l}},\ldots,e^{-j2\pi(M-1)\frac{d}{\lambda}sin\varphi_{k,l}}]$, where $\varphi_{k,l}=\arcsin(\frac\lambda d\phi_{k,l})$ is the AOA of the $l$th path of the user $k$ with the maximum angular spread $\delta=15^\circ$, the antenna spacing $d$ is set as half-wavelength. Note that the model contains complex variables, so we use the following formula to convert them to real variables to simplify the algorithm implementation process, which does not have any performance disadvantage.
\begin{equation}
    \begin{bmatrix}Re(\mathbf{\hat{Y}})\\Im(\mathbf{\hat{Y}})\end{bmatrix}=\begin{bmatrix}Re(\mathbf{\hat{S}})&-Im(\mathbf{\hat{S}})\\Im(\mathbf{\hat{S}})&Re(\mathbf{\hat{S}})\end{bmatrix}\begin{bmatrix}Re(\mathbf{\hat{X}})\\Im(\mathbf{\hat{X}})\end{bmatrix},
\end{equation}
In fact, our proposed CE-SBL algorithm realizes the joint estimation of the two parts by making the real and imaginary parts of $\mathbf{\hat{X}}$ share the same hyperparameter set. For the specific setting of the hyperparameter contained in the algorithm, the numerical experiments show that our algorithm is not strict on hyperparameters. As long as it is in a reasonable region $a\in(25,50)$, $c\in(0,1)$, our algorithm can achieve ideal performance. In the experiment below, we set $a=30$, $c=0.125$. Additionally, to better align with the cluster-extended structure of the sparse signal, we set the trade-off parameter $\lambda=0.01$ and the correlation parameter $\kappa=0.1$, enabling the CE-SBL algorithm to achieve optimal performance.
\par
As a reference, we compare the proposed CE-SBL algorithm with existing advanced algorithms, which include i) Turbo Generalized MMV AMP (Turbo-GMMV-AMP)\cite{ke2020compressive}, ii) AMP\cite{liu2018massive}, iii) SBL\cite{wipf2007empirical}, iv) Sparsity Adaptive Matching Pursuit (SAMP)\cite{do2008sparsity}, and v) $\ell$2 reweighting algorithm\cite{chen2021joint}. 
\par
As a performance measure for AD and CE, we use the following two indicators as well:
\begin{itemize}
    \item The active user detection ratio $\mu_{AD}$ for AD: 
    \begin{equation}
        \mu_{AD}=\frac{K_{AD}}{K_{all}},
    \end{equation}
    where $\hat{K}_{AD}$ and $\hat{K}_{all}$ represent the number of successfully detected active users and all active users respectively.
    \item The NMSE of CE, which is defined as:
    \begin{equation}
        \text{NMSE}_{CE}=10\log_{10}\frac{\|\mathbf{\hat{X}}_{CE}-\mathbf{\hat{X}}_{act}\|_{\mathrm{F}}^2}{\|\mathbf{\hat{X}}_{CE}\|_{\mathrm{F}}^2},
    \end{equation}
    where $\mathbf{\hat{X}_{CE}}$ and $\mathbf{X}_{act}$ represent the traditional CSI matrix and the delay-angle domain CSI matrix, respectively.
\end{itemize}
\par
\begin{figure}[!t]
\centering
\includegraphics[width=3.45in]{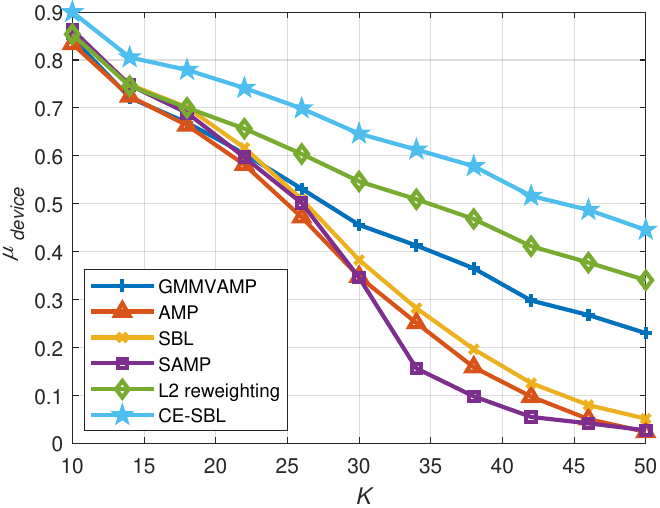}
\caption{Comparison of the active user detection ratio across a varying number of active users $k$.}
\label{1.1}
\end{figure}

\begin{figure}[!t]
\centering
\includegraphics[width=3.45in]{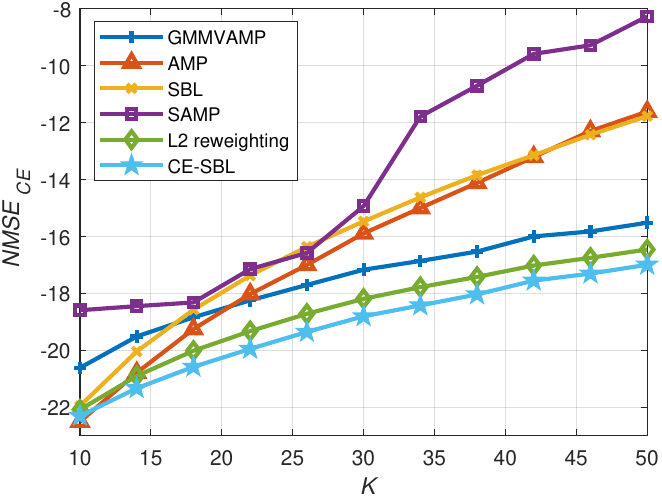}
\caption{Comparison of the NMSE of CE across a varying number of active users $k$.}
\label{1.2}
\end{figure}
Note that the number of successfully detected users, $K_{AD}$, and their respective delays, $t_k$, cannot be obtained directly, as the framework and algorithm are implemented in the delay-angle domain. We determine $K_{AD}$ and $t_k$ using the following approach: First, we sort the rows of the CSI matrix $\hat{X}_{CE}$ by their energy levels and then select the minimum number of rows whose combined energy exceeds $\theta_1\%$ of the total energy. The pilots corresponding to these selected rows are considered detected. In the experiments, we set $\theta_1\% = P_S / (P_N + P_S) = (1 + 10^{(-\text{SNR}/10)})^{-1}$, where $P_S$ and $P_N$ represent the power of the original signal and noise, respectively. An increase in $\theta_1$ raises the probability of false detection and lowers the probability of missed detection, while a decrease in $\theta_1$ has the opposite effect. Next, for each detected row of $X_{CE}$, we select the minimum number of elements whose combined power exceeds $\theta_2$ of the total power of the row. Finally, if two adjacent elements in any row exceed $\theta_3$, a cluster is considered to have formed in that row, indicating multiple active users. In other words, if the AOAs of two users are separated more than $180^\circ\times\theta_3/M$, the two users can be differentiated in the angular domain even if they select the same pilot. In the experiment, $\theta_2$ and $\theta_3$ are set to 0.98 and 3, respectively. Distinct clusters in each row of $X_{CE}$ correspond to different active users that have selected the same pilot. Accordingly, based on the transformation in (\ref{eq3}), we recover the channel coefficient vector of each active user sequentially by retaining only one cluster in the vector $\hat{\mathbf{x}}_i$ and setting all other clusters to zero. An active user is considered successfully detected only if its delay estimation is correct and the normalized mean square error (NMSE) of the estimated channel coefficient vector $\mathbf{\hat{d}}_{CE}$ meets the following criterion:
\begin{equation}
    \text{NMSE}_{\mathbf{\hat{d}}_{CE}}=10\log_{10}\frac{\|\mathbf{\hat{d}}_{CE}-\mathbf{\hat{d}}_{act}\|_{\mathrm{F}}^2}{\|\mathbf{\hat{d}}_{CE}\|_{\mathrm{F}}^2}\leq-15\text{dB},
\end{equation}
where $\mathbf{\hat{d}}_{act}$ denotes the actual channel coefficient vector. And the delay $t_k$ of active user $k$ can be easily calculated by taking the remainder of the maximum delay $t_m$ from the index of their vector $\hat{\mathbf{x}}_i$ in the original CSI matrix $\mathbf{\hat{D}}$.
\par
Fig. \ref{1.1} and Fig. \ref{1.2} illustrate the performance of AD and CE, respectively, with varying numbers of active users $K$. It can be observed that no matter how the number of active users changes, our proposed method has better performance than other comparison methods. In all cases with more than 20 active users, the active user detection ratio $\mu_{AD}$ of the CE-SBL algorithm is at least 10\% higher than that of the best comparison algorithm, and channel estimation approaches near-perfect accuracy. All methods experience performance deterioration as the number of active devices increases. As the number of active users increases, the possibility of users choosing the same pilot frequency also increases, resulting in the increasing possibility of user conflicts. The AOA and delay information of users can help resolve the conflicts and distinguish the users who choose the same pilot. The performance of the proposed method improves as the number of antennas or the maximum system delay increases, providing greater dimensionality to distinguish conflicting users. However, these adjustments would also increase computational complexity.
\par
\begin{figure}[!t]
\centering
\includegraphics[width=3.45in]{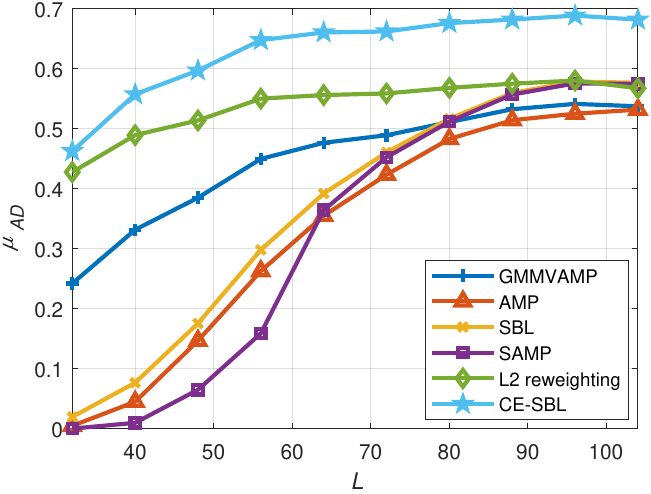}
\caption{Comparison of the active user detection ratio across a varying pilot length $L$.}
\label{2.1}
\end{figure}

\begin{figure}[!t]
\centering
\includegraphics[width=3.45in]{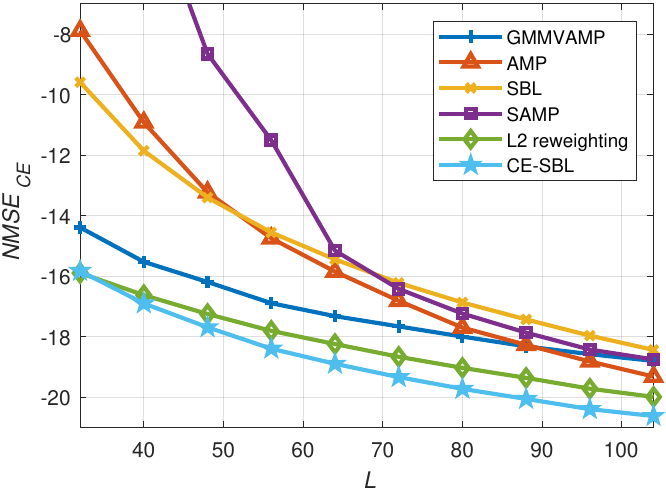}
\caption{Comparison of the NMSE of CE across a varying pilot length $L$.}
\label{2.2}
\end{figure}
Fig. \ref{2.1} and Fig. \ref{2.2} illustrate the performance of AD and CE, respectively, with varying pilot lengths $L$. It can be observed that as the pilot length increases, the performance of both indicators improves. Additionally, the CE-SBL algorithm outperforms all comparison algorithms. It can achieve comparable performance to other algorithms with shorter pilot lengths. For instance, state-of-the-art methods require a pilot length of at least $L=80$ to achieve an active device detection ratio $\mu_{device} > 0.6$, whereas the proposed method achieves this ratio with $L \leq 48$, indicating a 40\% reduction in spectrum resource usage. This gain results from the use of delay-angle information and an algorithmic design that leverages the cluster-extended structure.
\par
\begin{figure}[!t]
\centering
\includegraphics[width=3.45in]{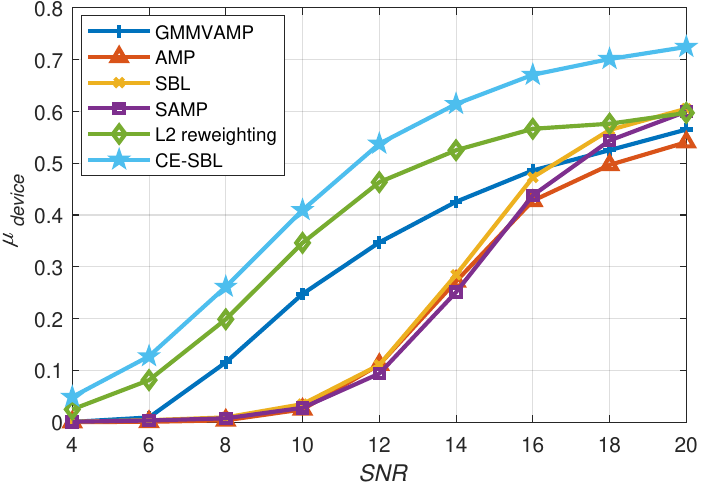}
\caption{Comparison of the active user detection ratio across a varying SNR.}
\label{3.1}
\end{figure}

\begin{figure}[!t]
\centering
\includegraphics[width=3.45in]{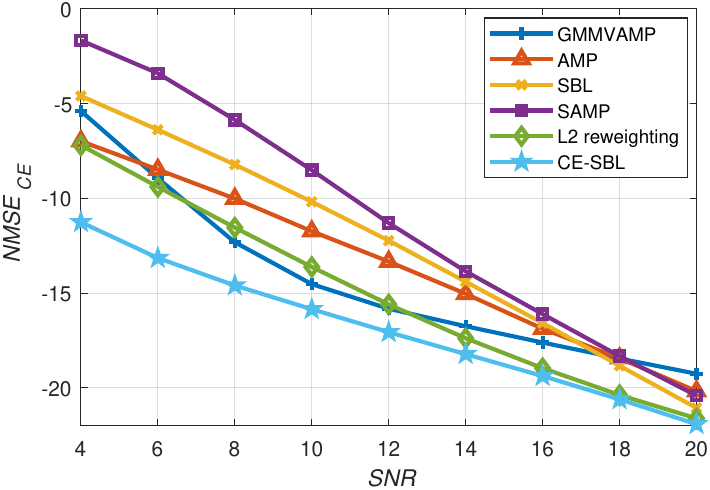}
\caption{Comparison of the NMSE of CE across a varying SNR.}
\label{3.2}
\end{figure}
Fig. \ref{3.1} and Fig. \ref{3.2} illustrate the performance of AD and CE, respectively, with varying SNR. In terms of active user detection $\mu_{device}$, the comparison algorithm can achieve a successful detection rate of up to 60\%, while the proposed method can achieve at least 72\%. Additionally, the larger the SNR is, the performance gain of the proposed method increases from 2\% to 12\%, and the performance gain is more obvious. This shows the superior performance of our algorithm in resolving user contention. In terms of channel estimation $\text{NMSE}_{CE}$, The proposed method outperforms all comparison methods, especially in the case of low SNR, which can improve $\text{NMSE}_{CE}$ by up to 5$\text{dB}$. This shows that the method has better robustness and can maintain an excellent level under a complex environment. 
\par
\begin{figure}[!t]
\centering
\includegraphics[width=3.45in]{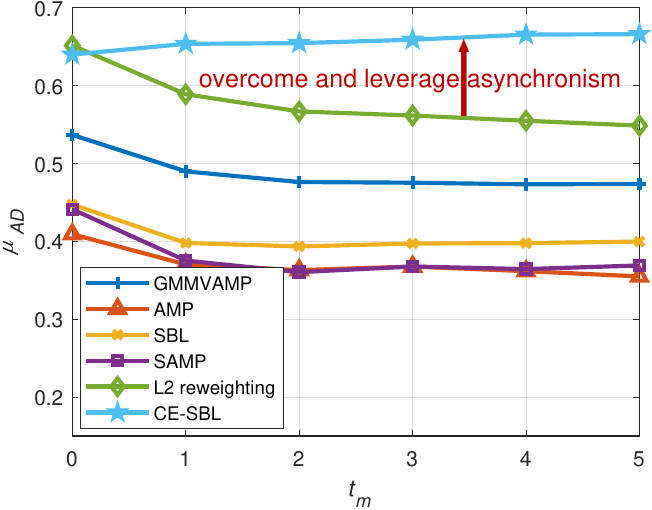}
\caption{Comparison of the active user detection ratio across a maximum delay $t_m$.}
\label{4.1}
\end{figure}

\begin{figure}[!t]
\centering
\includegraphics[width=3.45in]{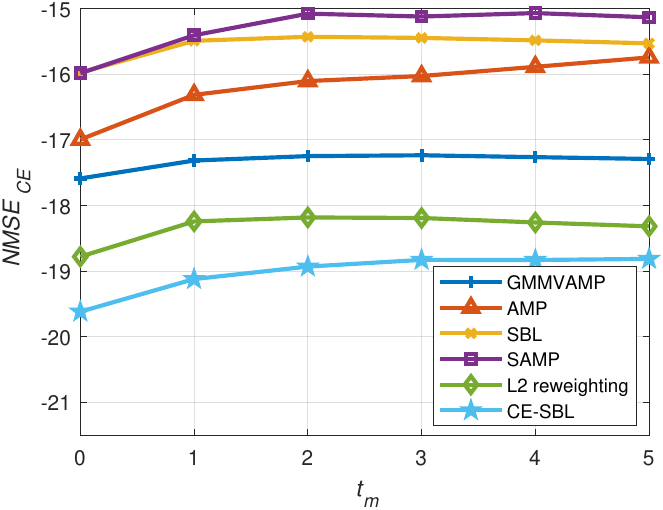}
\caption{Comparison of the NMSE of CE across a maximum delay $t_m$.}
\label{4.2}
\end{figure}
Finally, it is instructive to measure the required level of asynchronism that brings benefits in terms of access. We measure the strength of asynchronism through the variance of active user delay. Specifically, we adjust the maximum system delay $t_m$ to reflect the variance across all the active user delays. Fig. \ref{4.1} and Fig. \ref{4.2} illustrate the performance of AD and CE, respectively, with varying maximum delay $t_m$. When $t_m=0$, it can be seen as a synchronous system. The proposed algorithm does not provide advantages and the performance of it is close to that of the $\ell2$ reweighting algorithm in AD. When $t_m>0$, all competing algorithms exhibit performance, while the performance of the proposed algorithm improves due to the advantageous use of asynchronism. Additionally, since we have solved the contention of users selecting the same pilots when the variance is small, there is little change in AD and CE when the maximum user delay increases, hereby indicating the robustness of the algorithm. Even under excessive asynchronism, the proposed algorithm can still maintain the performance. Finally, in terms of CE, the proposed algorithm performs well since it is designed for specific cluster-extended signal structures.
\section{Conclusion}
This paper examines asynchronous grant-free CBRA in massive MIMO systems. To improve the accuracy of massive random access, we address the delay issue caused by asynchronous user access and leverage it as valuable information for user detection. We incorporate the delay from asynchronous systems and the angle information from massive MIMO to enhance AD and CE performance, formulating the problem as a sparse recovery LIP in the delay-angle domain. The recovery matrix exhibits both row-sparse and cluster-sparse structures with unknown sizes and supports, presenting a challenge for signal recovery. To jointly perform AD and CE under these challenging conditions, we develop the cluster-extended sparse Bayesian learning (CE-SBL) algorithm, which introduces a new weighted prior to capture the signal structure and estimates the parameters iteratively. Additionally, we provide a theoretical analysis of the algorithm and derive parameter bounds to demonstrate the effectiveness of the proposed model and algorithm.


\appendix[Proof of Theorem 1]
We show that the maximum sparse solution $\hat{r}_m$ of $\mathbf{\hat{X}}$ is $\lfloor M/D\rfloor(\lceil(\hat{L}+D)/2\rceil-1)$ and $\hat{r}_m>r_m$. Given that the maximum length of a cluster is $D$, then at most $\lfloor M/D\rfloor$ clusters can be formed in a row. Let's start by assuming an ideal situation where the column indexes of all clusters are aligned. Let $\mathbf{\hat{Y}}=[\mathbf{\hat{Y}}_1,\ldots,\mathbf{\hat{Y}}_m]$ and $[\mathbf{\hat{X}}=[\mathbf{\hat{X}}_1,\ldots,\mathbf{\hat{X}}_m]$, where $m>\lfloor M/D\rfloor$, then the columns in which each cluster resides can be viewed as a small MMV problem. The original problem consisted of several such small MMV problems:
\begin{equation}
    \mathbf{\hat{Y}}_1=\mathbf{\hat{S}}\mathbf{\hat{X}}_1,\ldots,\mathbf{\hat{Y}}_m=\mathbf{\hat{S}}\mathbf{\hat{X}}_m.
\end{equation} 
The set consisting of the number of columns of $\mathbf{\hat{X}}_j$ is $D_{set}=\{d_1,\ldots,d_m\}$ with $d_j<D$. Hence, referring to Lemma 1, the maximum sparse solution that the $j$-th small MMV problem can support is $\lceil(\hat{L}+d_j)/2\rceil-1$. When we combine all these small MMV problems, at least the maximum sparse solution that the framework can support in total is:
\begin{equation}
    \hat{r}_m=\sum_{j=1}^m\left(\left\lceil\frac{\hat{L}+d_j}{2}\right\rceil-1\right).
\end{equation} 
When the length of all clusters is $D$, the maximum sparse solution reaches a minimum value:
\begin{equation}
    \min(\hat{r}_m)\geq\lfloor M/D\rfloor(\lceil(\hat{L}+D)/2\rceil-1).
\end{equation}

\par
Note that actually there can be no ideal alignment between clusters in different rows, but what we want to say is that if clusters in different rows are interleaved on the columns, then the maximum sparse solution supported will be larger because this can be seen as more small MMV problem than the ideal case with smaller dimensions. Therefore, we consider the worst problem, which forms the most clusters.
\par
Next, we show that $\hat{r}_m$ in our framework is greater than $r_m$ in the traditional framework. We calculate the residual difference between $\hat{r}_m$ and $r_m$:
\begin{equation}
\begin{aligned}
    \Delta&= \left\lfloor \frac{M}{D}\right\rfloor\left(\left\lceil\frac{\hat{L}+D}{2}\right\rceil-1\right)-\left(\left\lceil\frac{\hat{L}+M}{2}\right\rceil-1\right)\\
    &\geq \left(\frac{M}{D}-1\right)\left(\frac{\hat{L}+D}{2}-1\right)-\frac{\hat{L}+M}{2}\\
    &=\frac{M}{2D}(\hat{L}-2)+\frac{D}{2}+\hat{L}+1.
\end{aligned}
\end{equation}
Known that $L$, $M$, and $D$ are positive and $\hat{L}\gg2$, we get $\Delta>0$, thus $\hat{r}_m>r_m$. Now the proof is completed.

\bibliographystyle{IEEEtran}  
\bibliography{reference}


 





\end{document}